\documentclass[sageapa, times, Afour]{sagej}

\usepackage{moreverb,url}
\usepackage{subcaption}
\usepackage{multirow}
\usepackage{xurl}
\usepackage[colorlinks,bookmarksopen,bookmarksnumbered,citecolor=red,urlcolor=red]{hyperref}

\newcommand\BibTeX{{\rmfamily B\kern-.05em \textsc{i\kern-.025em b}\kern-.08em
T\kern-.1667em\lower.7ex\hbox{E}\kern-.125emX}}

\def\volumeyear{2024}

\begin{document}

\runninghead{Reeves et al.}

\title{Exploring the Impact of ChatGPT on Wikipedia Engagement}

\author{Neal Reeves, Wenjie Yin and Elena Simperl}

\affiliation{Department of Informatics, King's College London}

\corrauth{Neal Reeves, 
Bush House,
King's College London,
Strand,
London,
United Kingdom,
WC2R 2LS.}

\email{neal.t.reeves@kcl.ac.uk}

\begin{abstract}
Wikipedia is one of the most popular websites in the world, serving as a major source of information and learning resource for millions of users worldwide. While motivations for its usage vary, prior research suggests shallow information gathering -- looking up facts and information or answering questions -- dominates over more in-depth usage. On the 22nd of November 2022, ChatGPT was released to the public and has quickly become a popular source of information, serving as an effective question-answering and knowledge gathering resource. Early indications have suggested that it may be drawing users away from traditional question answering services such as Stack Overflow, raising the question of how it may have impacted Wikipedia. In this paper, we explore Wikipedia user metrics across four areas: page views, unique visitor numbers, edit counts and editor numbers within twelve language instances of Wikipedia. We perform pairwise comparisons of these metrics before and after the release of ChatGPT and implement a panel regression model to observe and quantify longer-term trends. We find no evidence of a fall in engagement across any of the four metrics, instead observing that page views and visitor numbers increased in the period following ChatGPT's launch. However, we observe a lower increase in languages where ChatGPT was available than in languages where it was not, which may suggest ChatGPT's availability limited growth in those languages. Our results contribute to the understanding of how emerging generative AI tools are disrupting the Web ecosystem.
\end{abstract}

\keywords{Collective Intelligence, Wikipedia, ChatGPT, Engagement, Open Collaboration}

\maketitle

\section{Introduction}

Wikipedia is the largest online crowdsourced encyclopedia, consisting of over 6.6 million articles in 292 languages as of 2023 \cite{ren2023did}. While visitors cite a range of motives, studies have demonstrated fact lookup and fact checking as common factors that drive readers to consult articles \cite{lemmerich2019world,singer2017we}. Wikipedia is a crucial learning resource for US adults \cite{kross2021characterizing} and is commonly used -- albeit not necessarily effectively -- in areas such as health \cite{huisman2021health,smith2020situating}, programming \cite{robillard2020understanding} and history \cite{singer2017we} to name but a few. Prior work has also characterised Wikipedia as a "gateway to the Web", providing essential context to search engine results and bridging the gap to a range of external resources \cite{piccardi2021value}. Indeed, Wikipedia results appear highly frequently in important, trending searches, particularly on desktop devices \cite{vincent2021deeper}. Nevertheless, in spite of this characterisation, evidence suggests that users will often rely heavily on Wikipedia results for information needs, only proceeding to external sources and following citations when met with low quality content or when otherwise unable to meet those needs \cite{piccardi2020quantifying,piccardi2023large}. 

Wikipedia and its user base rely significantly on artificial intelligence tools to maintain and manage content. A substantial portion of edits within the platform are performed automatically or semi-automatically by user-maintained bots \cite{zheng2019roles}. Issues associated with the use and the community's perceptions of AI within Wikipedia are not universally positive with community's sometimes actively resisting or otherwise limiting the adoption of AI technologies in Wikipedia \cite{halfaker2020ores,smith2020keeping,teblunthuis2021effects}. Nevertheless, questions regarding the community's perceptions and use of AI tools are a current area of research focus and there is limited evidence of how existing tools and systems have influenced the community\endnote{\url{https://meta.wikimedia.org/wiki/Artificial_intelligence/Bellagio_2024} [Retrieved May 2024]}.  

Arguably one of the most disruptive AI tools of recent years has been ChatGPT. Launched in November 2022, the AI-powered chat-bot had reached 100 million users within under 3 months making it the ``fastest growing consumer internet app" ever\endnote{\url{https://www.theguardian.com/technology/2023/feb/02/chatgpt-100-million-users-open-ai-fastest-growing-app} [Retrieved May 2024]}. One key advantage linked to its early adoption is its ability to quickly and accurately summarise information to support searches \cite{ray2023chatgpt}. ChatGPT is already able to outperform search engines in some knowledge gathering tasks \cite{xu2023chatgpt}. Early indications suggest that it can give satisfactory answers to general and cultural fact-finding questions assuming they do not concern concepts more recent than its training data (i.e., 2021) \cite{amaro2023ai}. Nevertheless, results are prone to hallucinations, partial-truths or misinformation, particularly with attribution or when asked about controversial topics \cite{mcintosh2023culturally,zuccon2023chatgpt}. 

Question-answering platforms may already be impacted by ChatGPT. In the online crowdsourced question and answer service Stack Overflow, there was a 16\% reduction in the number of English language questions asked following the launch of ChatGPT, an effect not seen in Russian and Chinese where ChatGPT was unavailable \cite{del2023large}. Similar effects have been seen in the Stack Exchange platform \cite{sanatizadeh2023information}. Nevertheless, this reduction in engagement on Stack Exchange has also coincided with a general increase in the quality and complexity of the questions asked by users \cite{sanatizadeh2023information}. Previous research has demonstrated that these platforms rely heavily on Wikipedia to provide additional context for answers and comments \cite{baltes2020contextual} although the impact on Wikipedia editing and viewing behaviour has generally been limited \cite{vincent2018examining}. It also stands to reason that there may be some overlap between the capabilities of ChatGPT and Wikipedia as GPT-3 was itself trained in part on Wikipedia \cite{brown2020language}. A 2023 article published in the New York Times cited an anonymous Wikipedia editor's fears that generative AI tools may lead to the death of Wikipedia citing this overlap in capabilities as potential evidence \endnote{\url{https://www.nytimes.com/2023/07/18/magazine/wikipedia-ai-chatgpt.html} [Retrieved May 2024]}.

 Given these potential and observed impacts on resources which link to and use Wikipedia, we ask whether the rise of ChatGPT has had an impact on Wikipedia usage patterns. If Web users are adopting other sources of information, are these replacing or supplementing existing Web sources such as Wikipedia? And how does this vary between highly-resourced and less highly-resourced languages? We analyse usage statistics from Wikipedia between 2021 and 2024 to explore any impact of the launch of ChatGPT's public service on engagement in Wikipedia. We explore and compare twelve languages including six from countries where ChatGPT was unavailable and six selected based on the size and popularity of their Wikipedia resources and their representation in the training data used by ChatGPT to explore whether highly-resourced languages were more or less impacted than less represented and resourced languages. We report aggregate comparisons across four metrics before performing a pairwise comparison of engagement in the year prior to and following the launch of ChatGPT. We then perform a panel regression to explore, quantify and compare longer-term trends across the twelve languages. Our work provides evidence of the impact that LLMs and particularly ChatGPT have had on Wikipedia. 

\section{Background and Related Work}

\subsection{Wikipedia engagement and events}

Existing evidence demonstrates the impact that internal and external events can have on how users engage with Wikipedia. Media reports around particular scientific and technology-related topics, natural disasters and health-related crises have been demonstrated to drive engagement with Wikipedia \cite{segev2017temporal}. Similarly, region-specific political events are associated with spikes in page-views in relevant language editions of Wikipedia, such as the Kashmir Solidarity day leading to increased engagement within Urdu Wikipedia. This may even stretch beyond pages specific to those topics, with evidence of increased cooperation and engagement in various pages covering historical events during the Black Lives Matter social movement even where there was little to no new information around those events \cite{twyman2017black}. Seasonal events and phenomena such as the behaviours of animals have been shown to align with seasonal patterns of interest in particular Wikipedia pages, particularly for insects and flowering plants \cite{mittermeier2019season}. Technological changes can also have significant and pervasive changes in user behaviour as demonstrated by the significant and persistent drop in page-views observed in 2014 when Wikipedia introduced a page preview feature allowing desktop users to explore Wikipedia content without following links \cite{chelsy2019detecting}.

\subsection{Wikipedia as multilingual platform}

Given Wikipedia's status as a platform with a high number of articles across a wealth of languages, we are far from the first to compare activity across languages. Miz et al., analysed topics in the most popular trending articles across the English, French and Russian Wikipedia, finding that while there were some variations in terms of local events and cultural issues, trending topics were broadly similar across all three languages \cite{miz2020trending}. Nonetheless, trends in pageview statistics are unique to a given country and language edition, as shown by Xie et al. \cite{chelsy2019detecting}. Furthermore, when it comes to controversial topics, discrepancy in tone, organisation, and information presented can exist in different language editions. For example, such a discrepancy was seen between English, Hindi, and Urdu on the political conflict in Kashmir and Jammu \cite{hickman2021understanding}. Kub\'s similarly demonstrates the presence of culture-specific viewpoints in different language editions of articles on historical events \cite{kubs2021historical}. Miquel-Rib\'e and Laniado explored the growth and cross-cultural context of 40 different language Wikipedia editions with the authors exploring the Cultural Context Content (geography, people, language, traditions, etc.) associated with each language  \cite{miquel2018wikipedia}. Results demonstrated that the growth of individual languages was highly variable, but also that much of the cross-cultural context of each language is recorded uniquely in that language. Even in English, which tended to cover more of the Cultural Context Content from other languages, the percentage coverage was on average only ~34\%. Lemmerich et al., analysed motivations for reading Wikipedia across 14 language editions, finding that Wikipedia browsing behaviours were associated with the socio-economic properties of the relevant countries \cite{lemmerich2019world}. Inter-language diversity is not limited only to textual material with images showing diversity exceeding that of textual material even though images likely do not require translation \cite{he2018the_tower_of_babel}. Differences between languages also extend beyond simply content. The balance of visitors of each gender differs quite significantly across different language editions of Wikipedia and evidence suggests gendered distinctions in behaviour -- for example, that women view fewer pages in a session than men \cite{johnson2021global}. 

\subsection{Artificial intelligence and Wikipedia}

Much of the research exploring the role and impact of artificial intelligence and automated systems on Wikipedia has focused on introducing such tools to the platform and community. The use of machine agents within Wikipedia itself is nothing new as bots play a range of roles in creating and maintaining articles, particularly making fixes to page content \cite{zheng2019roles}. The relatively common use of bots is observed across languages although bot behaviours and applications are language and culture dependent \cite{tsvetkova2017even}. Bots play a crucial role in facilitating certain types of Wikipedia activities among more experienced moderators and there exists a formal approval process with opportunities for the community to discuss individual bots and their applications \cite{geiger2017beyond}. Despite these existing roles, however, the reception that AI tools have received has not been universally positive. A study of perceptions of the use of AI among the Wikipedia community identified five key values associated with its use, most notably a desire that humans have the final say in any decision making \cite{smith2020keeping}.

Even so, there is evidence to suggest positive reception to the use of AI in Wikipedia. Rececent research has shown that when presented with alternative citations for poorly supported claims generated by an algorithmic agent, Wikipedia users were twice as likely to prefer the algorithmic alternative \cite{petroni2023improving}. Additionally, the ORES system aims to algorithmically quality assess Wikipedia articles and has elicited a range of responses from users but crucially has been adopted by Wikipedia editors across languages with minor adjustment \cite{halfaker2020ores}. One question we aim to explore is not whether the Wikipedia editors might use emergent LLM models, but rather how this might have influenced -- and continue to influence -- their engagement. 

\subsection{ChatGPT as complementary and replacement tool}

Within the contemporary literature, an emerging focus of research is the evaluation of the performance of ChatGPT and its underlying models on common tasks across specific domains. Koco\'n et al., analysed a range of NLP tasks on ChatGPT and compared performance with existing State of the Art models\cite{kocon2023chatgpt}. The authors found ChatGPT generally performed adequately, but was unable to match the performance of the models and this loss in performance was worse with more difficult and/or subjective tasks. Frieder et al., explored the performance of ChatGPT with mathematical tasks finding that it performs acceptably for undergraduate-level but not graduate-level mathematical tasks with the authors noting that the model performed worse than any graduate-level mathematical student would \cite{frieder2024mathematical}. Within the domain of healthcare, current studies suggest performance is generally moderate at best \cite{li2024chatgpt}, although ChatGPT has achieved a passing score on both the German and US medical licensing examinations \cite{gilson2023does,jung2023chatgpt}. Within finance, ChatGPT outperforms State of the Art models in sentiment analysis tasks by as much as 35\% \cite{fatouros2023transforming}. Similarly in programming, ChatGPT outperforms State of the Art code refinement algorithms although there remain some weaknesses in its approach \cite{guo2024exploring}. 

More relevant to our research, Bang et al., analysed the performance of ChatGPT across a range of languages and a range of tasks \cite{bang-etal-2023-multitask}. Their analysis demonstrated that ChatGPT performed less effectively with low-resourced languages, but particularly when it came to identify and generate text using languages with non-Latin scripts, even for medium or high-resourced languages \cite{bang-etal-2023-multitask}. Zhang et al., found a lower response accuracy for tasks in languages other than English in line with systems that convert text to and from English, which the authors suggested may be due to the ChatGPT training dataset consisting predominantly of monolingual English inputs \cite{zhang2023don}. Zhang et al., performed an analysis of the performance of a number of LLMs including GPT-5 and found that they continued to perform poorly with multilingual text, particularly when the languages in question are low-resource and/or use non-Latin scripts \cite{zhang2024m3exam}. We take these findings into account when choosing the languages used for our analysis.

\section{Data and Methods}

We gathered data for twelve languages from the Wikipedia API covering a period of twenty two months between the 1st of January 2021 and the 1st of January 2024. This includes a period of approximately one year following the date on which ChatGPT was initially released on the 30th of November 2022.

\subsection{Languages}

We decided to conduct our analysis using Wikipedia articles covering a range of languages selected to ensure geographic diversity covering both the global north and south. When selecting languages, we looked at three key factors:

\begin{enumerate}
    \item The common crawl size of the GPT-3 main training data as a proxy for the effectiveness of ChatGPT in that language.
    \item The number of Wikipedia articles in that language\endnote{Prior research has used the number of Wikipedia articles to identify whether a language is high, middle or low-resource \cite{robinson2023chatgpt}}.
    \item The number of global first and second language speakers of that language.
\end{enumerate}

We aimed to contrast languages with differing numbers of global speakers and languages with differing numbers of Wikipedia articles. We split the available languages into three categories based on the relative number of speakers, the number of Wikipedia articles for that language and the common crawl size and used this to select the most suitable languages. Notably, English has the highest number of Wikipedia articles, the largest common crawl size within the GPT-3 training data and among the largest number of first and second language speakers. We then selected the remaining languages from the categories to ensure as much diversity as possible:

\begin{itemize}
    \item Urdu and Swahili -- languages with high populations but low numbers of Wikipedia articles and common crawl sizes.
    \item Arabic -- a language with a medium sized population and a medium number of Wikipedia articles.
    \item Italian and Swedish -- languages with low populations but high numbers of Wikipedia articles.
\end{itemize}

We note that the common crawl size for all of these languages other than English was extremely small as a percentage of the overall training data. Additionally, at the time of writing ChatGPT officially supports Arabic, English, Italian and Urdu, but not support Swedish or Swahili although this appears to only extend to the language of the interface rather than any functionality within the model itself.

As a comparison, we also analysed six languages selected from countries where ChatGPT is banned, restricted or otherwise unavailable. The six languages selected are:
\begin{itemize}
    \item Amharic (spoken in Ethiopia)
    \item Farsi (spoken in Iran)
    \item Russian
    \item Tigrinya (spoken in Eritrea)
    \item Uzbek
    \item Vietnamese
\end{itemize}

We specifically selected these six as languages where first language speakers would be concentrated predominantly in the country where access was banned\endnote{Russian is a notable exception to this. However, ChatGPT remains banned in both Russia and Belarus and restricted in Ukraine, which represent the three languages where the vast majority of Russian speakers reside.}. Eritrea, Ethiopia, Uzbekistan and Vietnam did gain access to ChatGPT in November 2023 but were unable to officially access ChatGPT prior to this date. We include them in our sample due to the long period where ChatGPT was not available and the relatively small number of languages spoken predominantly in countries where ChatGPT was banned which still had access to Wikipedia. We also recognise these restrictions may not necessarily prevent speakers of these languages from accessing ChatGPT as they may use a VPN or gain access from a country without restrictions, but this would still serve as a barrier to access for many citizens of these countries. 

\subsection{Metrics}

In gathering data from the API, we focused on four metrics: page views, visitors, edits and editors. Page views are broadly analogous to the number of times a page has been visited, while edits represent a single update to a page as made by a given user. The size of this edit may be highly variable and while Wikipedia has policies governing edits, it is largely up to a user how large an edit might be. We do not attempt to quantitatively measure the size or complexity of an edit due in large part to the difficulty associated with monitoring how much of a given change a user has made. For example, if a user reverts a previous edit but does not correctly record this in the edit description, their edit may seem very complex despite the fact that the user has not made any active changes to the page. Moreover, it should be noted that page views do not necessarily reflect a separate visit to the page as edits are also recorded as page views. 

\subsection{Panel Regression}

To quantify and statistically analyse the size of any effect ChatGPT may have had on each of the gathered metrics, we use a panel regression method with fixed effects. For this, we use the following formula:

$y_{it} = \alpha_i +  \beta_1 \times launch_{it} + \beta_2 d_{i} + \beta_3 w_{i} + \beta_4 \times \textit{t}_{i} + \epsilon_{it}$

Where $\alpha$ represents a language-specific fixed effect, $\beta_1$ is the coefficient of interest representing the change in engagement in a given language platform during the period where ChatGPT had launched, $\beta_2$ and $\beta_3$ are seasonal effects associated with a given day of the week d and week of the year w respectively on engagement in that language, $\beta_4$ represents an effect associated with time in days \textit{t} in that language and $\epsilon$ represents the error term.

We include both a fixed effect and interaction with seasonality for each language, as prior studies have shown that trends in patterns of Wikipedia pageviews are language specific language \cite{chelsy2019detecting}. 

\subsection{Standardisation}

Each of the twelve languages in our sample has a unique pattern of participation and making comparisons within and between languages is difficult due to the large number of outliers and significant difference from one language to another. The level of engagement with Wikipedia in English, for example, is much greater than engagement with Wikipedia in any of the other 11 languages. Moreover, each language exhibits significant variation from one day to the next. To account for this, we used the inverse hyperbolic sine function to transform and standardise the levels of engagement across the languages. The choice of this transformation was due to the ease of interpretation, with coefficients representing percentage changes, but also due to the transformation allowing for inputs to reach zero without the need for significant adjustment. We note that this transformation has found use in existing work in this space such as \cite{del2023large}.

\section{Results}

\begin{figure*}
    \centering
    \begin{subfigure}{0.75\textwidth}
        \includegraphics[width=\linewidth]{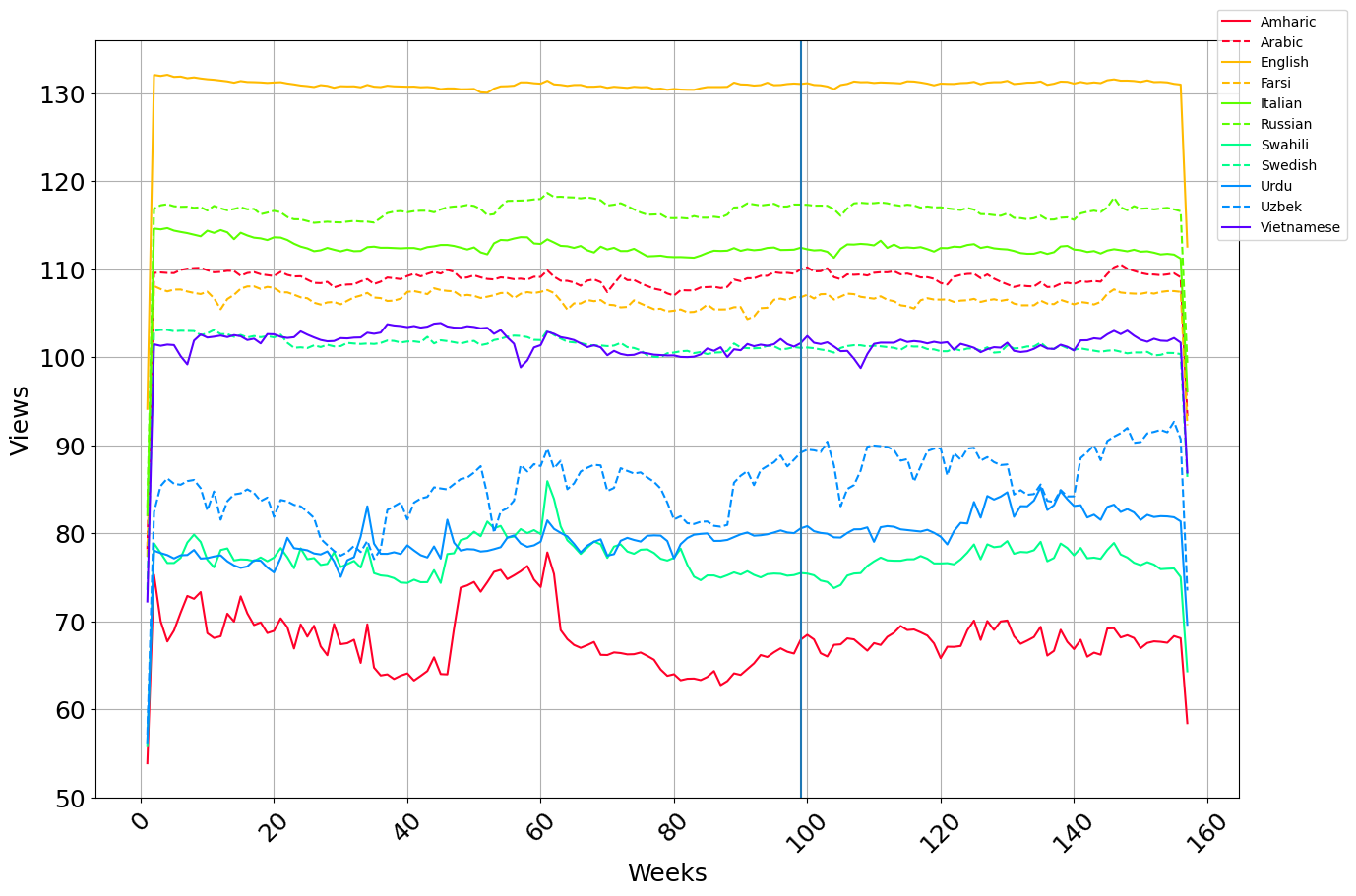}
    \end{subfigure}
    \begin{subfigure}{0.75\textwidth}
        \includegraphics[width=\linewidth]{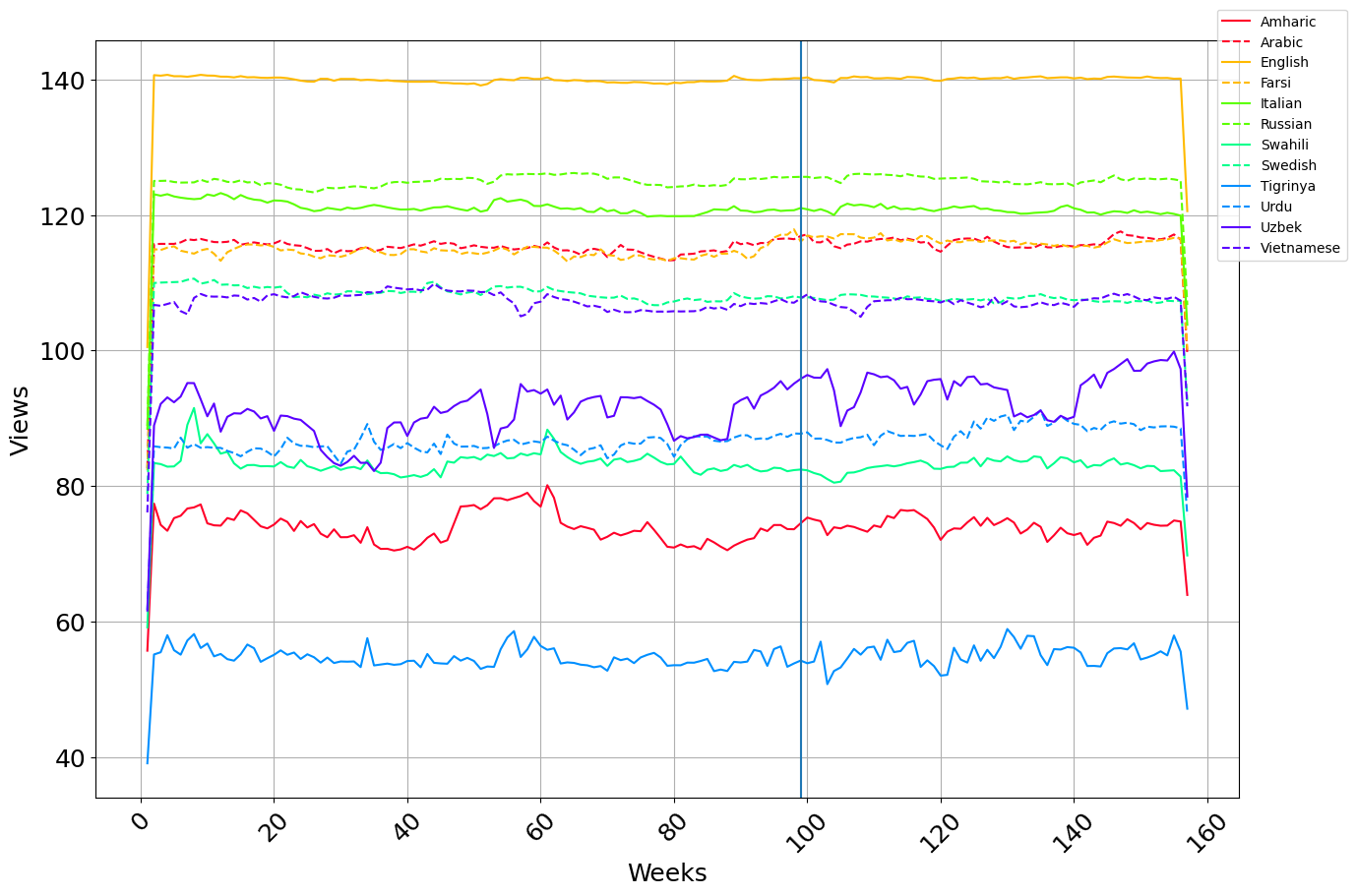}
    \end{subfigure}
    \caption{Standardised weekly total visitor (top) and page view (bottom) counts for each language.}
    \label{fig:twelve-visit}
\end{figure*}

\begin{figure*}
    \centering
    \begin{subfigure}{0.75\textwidth}
        \includegraphics[width=\linewidth]{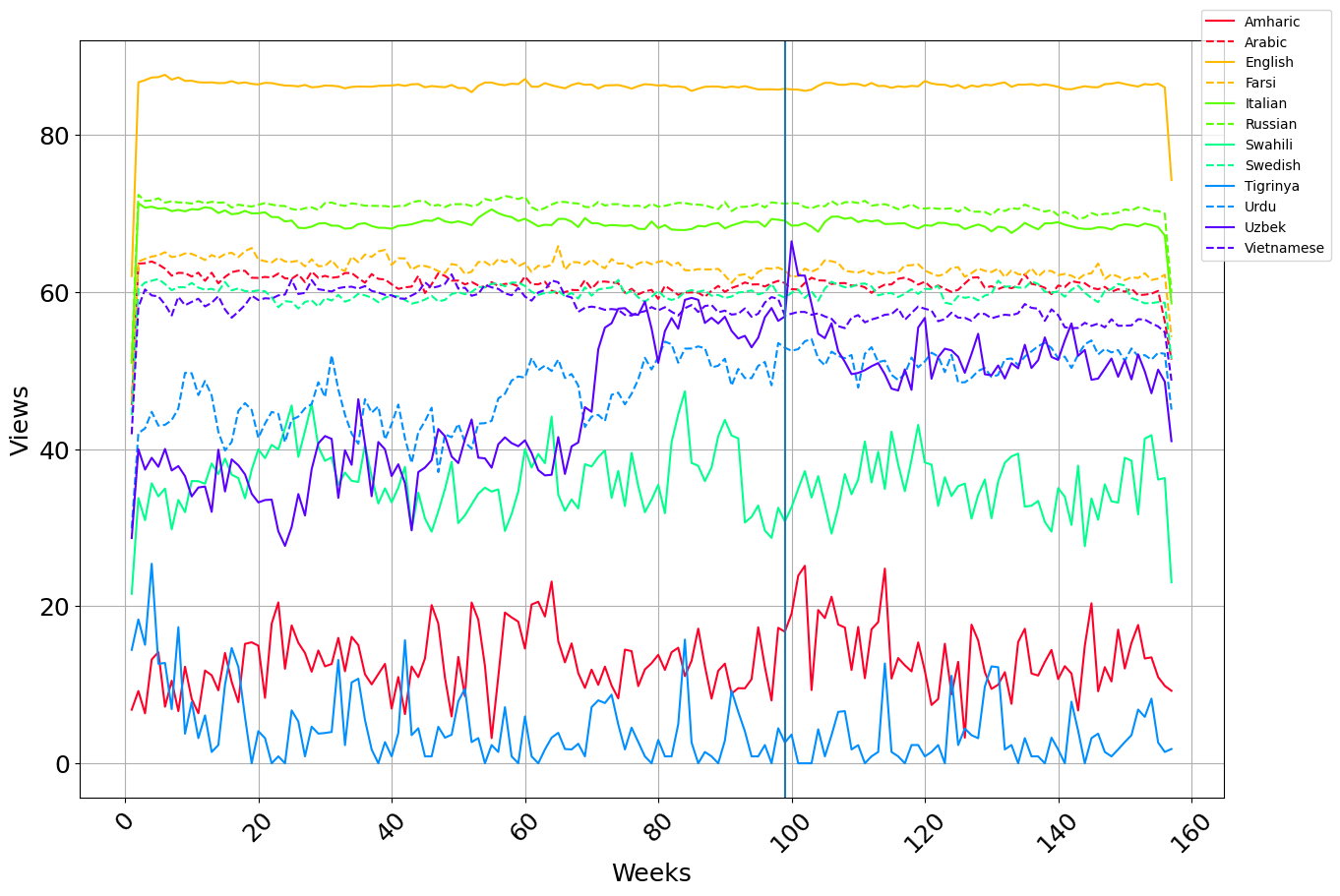}
    \end{subfigure}
    \begin{subfigure}{0.75\textwidth}
        \includegraphics[width=\linewidth]{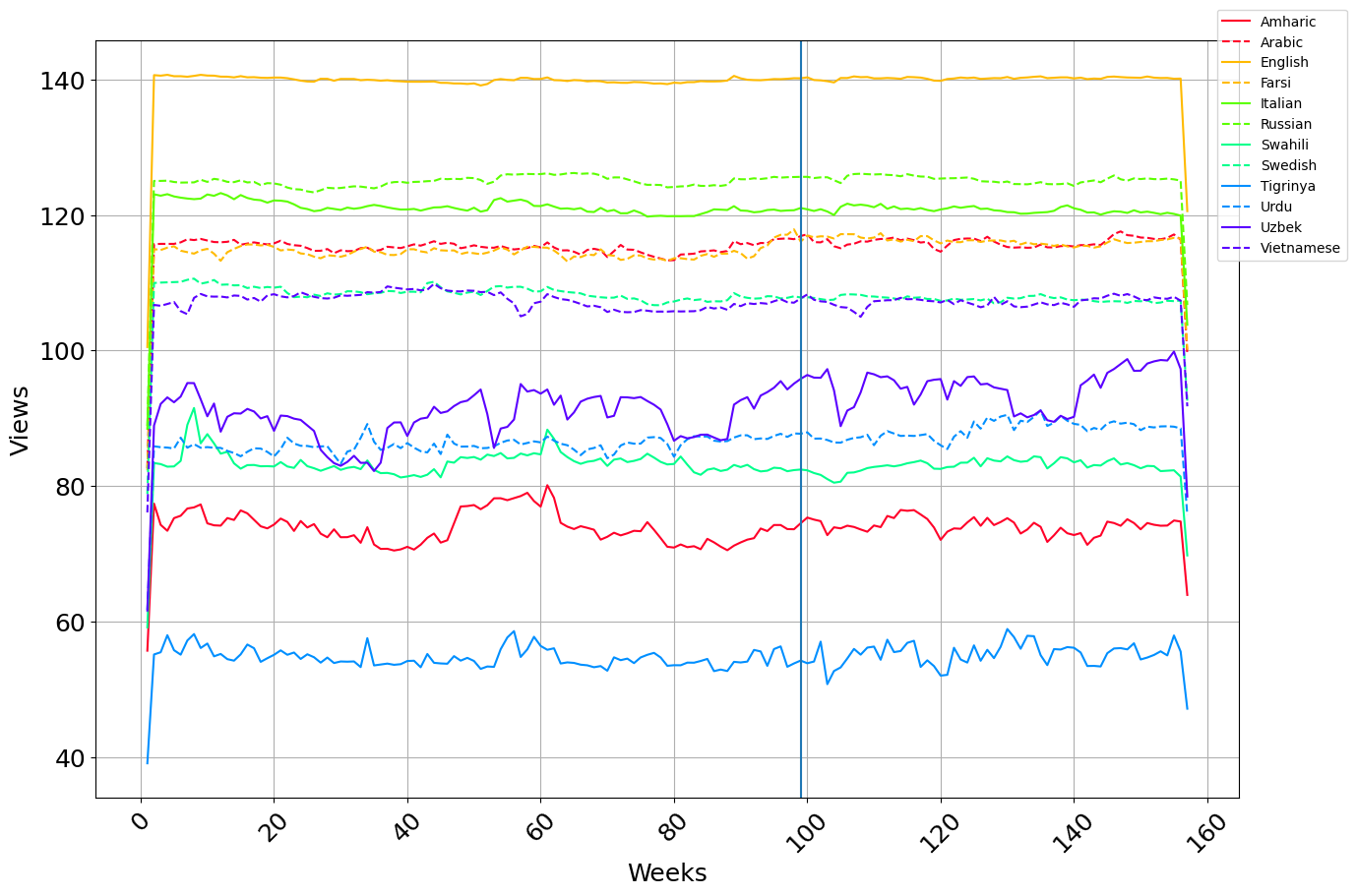}
    \end{subfigure}
    \caption{Standardised weekly total editor (top) and edit (bottom) counts for each language}
    \label{fig:twelve-edit}
\end{figure*}

We gathered data from the API covering a three year period from the 1st of January 2021 to the 1st of January 2024. Charts showing the standardised page views and visitor counts for each of the twelve languages can be seen in figure \ref{fig:twelve-visit} while editor and edit counts for can be seen in figure \ref{fig:twelve-edit}.

\subsection{Aggregate Comparisons}

As a first step to assess any impact from the release of ChatGPT, we performed paired statistical tests comparing aggregated statistics for each language for a period before and after release. We conducted this analysis for 4 engagement metrics: \textit{page views}, \textit{visitor numbers}, \textit{edits} and \textit{editors}. Due to the large number of languages involved in our analysis, we report all metrics with a Bonferroni correction.

\subsubsection{Page Views}

For page views, we first performed a two-sided Wilcoxon Rank Sum test to identify whether there was a difference between the two periods (regardless of directionality). We found a statistically significant different for five of the six languages where ChatGPT was available and two of the six languages where it was not. However, when repeating this test with a one-sided test to identify if views in the period after release were \textit{lower} than views in the period before release, we identified a statistically significant result in Swedish, but not for the remaining 11 languages.

\subsubsection{Visitors}

In contrast to page views, results for visitor numbers showed statistically significant results for a small number of languages and not all languages with a statistically significant different number in visitors demonstrated such a difference for page views. Among the languages where ChatGPT was available, we found statistically significant differences in four languages -- Arabic, English, Swedish and Urdu. Conversely, among the languages where ChatGPT was \textit{not} available, we identified statistically significant differences in only two languages -- Russian and Uzbek.  When performing the one-sided test, Swedish was once again the only language to show a statistically significant difference. 

\subsubsection{Edits}

In terms of edits, we observed statistically significant differences among five languages with four of these being languages where ChatGPT was unavailable -- Farsi, Russian, Uzbek and Vietnamese -- while two were languages where ChatGPT was available: Italian and Urdu. For the one-sided test, we found a difference only in four languages, three of which were languages where ChatGPT was unavailable (Farsi, Russian and Vietnamese) and only Italian among the languages where ChatGPT was available. 

\subsubsection{Editors}

For editors, there was no statistically significant difference between the period before and after the release of ChatGPT for most languages. We observed differences in only five languages -- Arabic and Italian for the languages where ChatGPT was available and Russian, Uzbek and Vietnamese for the languages where ChatGPT was not available. When performing the one-sided test, we observed statistically significant differences in Farsi, Italian, Russian and Vietnamese suggesting that all of these languages saw a statistically significant fall in edit numbers in the period after the release of ChatGPT.

\subsection{Panel Regression}

While the Wilcoxon Signed-Rank test provided weak evidence for changes among the languages before and after the release of ChatGPT, we note ambiguities in the findings and limited accounting for seasonality. To address this and better evaluate any impact, we performed a panel regression using data for each of the four metrics. Additionally, to account for longer-term trends, we expanded our sample period to cover a period of three years with data from the 1st of January in 2021 to the 1st of January 2024. Here, we assess the coefficient associated with each individual language and the coefficient associated with the interaction between the launch period and each language to assess the impact that the launch may have had.

\subsubsection{Page Views}

To better understand the effect that the availability of ChatGPT may have had on the number of page views in Wikipedia, we conducted a Panel Regression with each of the six languages. Results of this regression can be seen in Table \ref{tab:page-views}. For all six languages, we found a statistically significant difference in page views associated with whether ChatGPT had launched when controlling for day of the week and week of the year. In five of the six languages, this was a positive effect with Arabic featuring the most significant rise (18.3\%) and Swedish featuring the least (10.2\%). The only language where a fall was observed was Swahili, where page views fell by 8.5\% according to our model. However, Swahili page viewing habits were much more sporadic and prone to outliers perhaps due to the low number of visits involved.

We then analysed the six language versions of Wikipedia where ChatGPT is was unavailable. Once again, results showed a statistically significant rise across five of the six languages. However, in contrast with the six languages where ChatGPT was available, these rises were generally much more significant. For Farsi, for example, our model showed a 30.3\% rise, while for Uzbek and Vietnamese we found a 20.0\% and 20.7\% rise respectively. In fact, four of the languages showed higher rises than all of the languages where ChatGPT was available except Arabic, while one was higher than all languages except Arabic and Italian. 

\begin{table}
    \centering
    \begin{tabular}{cccc}
        Category & Language & Coefficient & p \\
        \hline
        \multirow{6}{*}{Available} & Arabic* & 0.1677 & \textless0.0001\\
        & English* & 0.0726 & \textless0.0001\\
        & Italian* & 0.1382 & \textless0.0001\\
        & Swahili* & -0.0888 & \textless0.0001\\
        & Swedish* & 0.0972 & \textless0.0001\\
        & Urdu* & 0.0816 & \textless0.0001\\
        \hline
        \multirow{6}{*}{Unavailable} & Amharic* & 0.1429 & \textless0.0001\\
        & Farsi* & 0.2649 & \textless0.0001\\
        & Russian & -0.0580 & 0.0752\\
        & Tigrinya* & 0.1216 & \textless0.0001\\
        & Uzbek* & 0.1825 & \textless0.0001\\
        & Vietnamese* & 0.1881 & \textless0.0001\\
        \hline
    \end{tabular}
    \caption{Panel regression result for page views. '= statistically significant at p\textless0.05, \textsuperscript{+}= significant at p\textless0.01, *= significant at p\textless0.001}.
    \label{tab:page-views}
\end{table}

\subsubsection{Edits}

Panel regression results for the six languages were generally not statistically significant. Among the languages where a significant result was found, our model suggested a 23.7\% rise in edits in Arabic, while for Urdu the model suggested a 21.8\% fall. We also observed a weakly statistically significant fall in Swahili of 9.8\%. For the highly resourced languages, English and Italian showed small rises and Swedish showed a very minor fall, but none of these were statistically significant. Similar findings were observed among the languages where ChatGPT was unavailable with a 12.9\% rise in Tigrinya and a 71\% fall in the case of Uzbek. However, as with Swahili, participation in the Uzbek Wikipedia was extremely sporadic and prone to variations. Other language results were not statistically significant. A summary of results can be seen in Table \ref{tab:edits}.

\begin{table}
    \centering
    \begin{tabular}{cccc}
        Category & Language & Coefficient & p \\
        \hline
         \multirow{6}{*}{Available} & Arabic\textsuperscript{+} & 0.2130 & 0.0032\\
        & English & 0.0506 & 0.4832\\
        & Italian & 0.0717 & 0.3202\\
        & Swahili & -0.1021 & 0.1572\\
        & Swedish & -0.0017 & 0.9812\\
        & Urdu* & -0.2458 & 0.0007\\
        \hline
        \multirow{6}{*}{Unavailable} & Amharic & 0.0725 & 0.3129\\
        & Farsi & 0.0301 & 0.6762\\
        & Russian & -0.0555 & 0.4418\\
        & Tigrinya* & 0.1216 & \textless0.0001\\
        & Uzbek* & -1.2466 & \textless0.0001\\
        & Vietnamese & -0.1138 & 0.1146\\
        \hline
    \end{tabular}
    \caption{Panel regression result for edits. '= statistically significant at p\textless0.05, \textsuperscript{+}= significant at p\textless0.01, *= significant at p\textless0.001}.
    \label{tab:edits}
\end{table}

\subsubsection{Visiting Users}

Wikipedia is a service that can be used by almost anyone provided they have an internet connection and are based in a country that does not ban access. There are no restrictions in terms of cost and a user does not need to have an account to view an article. As a result, the Wikimedia API does not provide any details on the number of users who visited pages on any given day, but it does provide the number of unique devices that accessed each language platform. We note that this metric is somewhat imperfect as an individual who uses more than one device would be counted multiple times, while multiple users who share a device would only be counted once. Additionally, a user who logs in only to edit would still be recorded as a visitor, but this is less of a concern as the Wikimedia API records any edit as a page view. Regardless, we chose to use the number of unique devices as the best available proxy for the number of visitors a page received. 

\begin{table}
    \centering
    \begin{tabular}{cccc}
        Category & Language & Coefficient & p \\
        \hline
        \multirow{6}{*}{Available} & Arabic* & 0.1442 & \textless0.0001\\
        & English' & 0.0765 & 0.0232\\
        & Italian* & 0.1133 & 0.0008\\
        & Swahili* & -0.1432 & \textless0.0001\\
        & Swedish' & 0.0670 & 0.0470\\
        & Urdu & 0.0610 & 0.0705\\
        \hline
        \multirow{6}{*}{Unavailable} & Amharic & 0.2944 & \textless0.0001\\
        & Farsi* & 0.1951 & \textless0.0001\\
        & Russian\textsuperscript{+} & -0.0994 & 0.0032\\
        & Tigrinya\textsuperscript{+} & 0.0993 & 0.0032\\
        & Uzbek' & 0.0783 & 0.0202\\
        & Vietnamese* & 0.1578 & \textless0.0001\\
        \hline
    \end{tabular}
    \caption{Panel regression result for visitor numbers. '= statistically significant at p\textless0.05, \textsuperscript{+}= significant at p\textless0.01, *= significant at p\textless0.001}.
    \label{tab:visitors}
\end{table}

As with the number of page views, the period after the release of ChatGPT was associated with an increase in visitor numbers in five of the languages according to our model as shown in Table \ref{tab:visitors}. We observe statistically significant increases for Arabic (15.5\%), English (8.0\%),  Italian (12.0\%), Swedish (6.9\%) and Urdu (6.3\%) with a statistically significant fall in Swahili (13.3\%). When compared with page views among the languages where ChatGPT was not available we were unable to explore values for Tigrinya where the API would only return data for a small number of days and so we did not include it in our model. Among the remaining languages, we once again observe a statistically significant rise among four of the languages and once again, these were generally higher than those for the languages where ChatGPT was available. The highest rise was observed in Amharic where our model predicted a rise of 29.6\%, followed by Farsi (21.4\%), Vietnamese (18.6\%) and Uzbek (10.1\%). We note these rises were larger than for all of the languages where ChatGPT was available with the exception of Uzbek. 

\subsubsection{Editors}

In addition to exploring the number of edits, we also analysed the number of editors that contributed to Wikipedia each week to explore whether there were changes in editor behaviours -- i.e., even though the number of edits may not have changed, this may have been due to a change in the number of edits each editor made. A summary of statistics for each language can be seen in Table \ref{tab:editors}.

Our model did not find a statistically significant change in editors for either English or Swedish. For Italian, we found a weakly statistically significant rise of 6.6\%, while for Urdu we found a weakly statistically significant fall of 6.7\%. The largest impact was observed in Arabic where the model identified a 12.5\% as associated with the period after ChatGPT released. Among the languages where ChatGPT was unavailable, we found a statistically significant rise in Farsi and Tigrinya but a statistically significant fall in Uzbek. 

\begin{table}
    \centering
    \begin{tabular}{cccc}
        Category & Language & Coefficient & p \\
        \hline
        \multirow{6}{*}{Available} & Arabic\textsuperscript{+} & 0.1173 & 0.0038\\
        & English & 0.0442 & 0.2746\\
        & Italian & 0.0638 & 0.1153\\
        & Swahili* & 0.1557 & 0.0001\\
        & Swedish & 0.0457 & 0.2593\\
        & Urdu & -0.0696 & 0.0856\\
        \hline
        \multirow{6}{*}{Unavailable} & Amharic & 0.0164 & 0.6860\\
        & Farsi* & 0.1414 & 0.0005\\
        & Russian & -0.0040 & 0.9204\\
        & Tigrinya\textsuperscript{+} & 0.1313 & 0.0012\\
        & Uzbek* & -0.6370 & \textless0.0001\\
        & Vietnamese & -0.0403 & 0.3200\\
        \hline
    \end{tabular}
    \caption{Panel regression result for editor numbers. '= statistically significant at p\textless0.05, \textsuperscript{+}= significant at p\textless0.01, *= significant at p\textless0.001}.
    \label{tab:editors}
\end{table}

\section{Discussion}

\subsection{Impact on Views}

Making conclusive statements about the role ChatGPT may have played in any changes within Wikipedia is inevitably difficult. Each of the languages we explored has a number of confounding factors associated with it and one or all of these factors may have played a significant role in the changes our model predicted. Nevertheless, far from being associated with \textit{reduced} engagement, our regression outputs actually suggest that for five of the six languages, the number of page views and the number of visiting users actually increased in the period after ChatGPT released -- arguably substantially so given that the smallest increase we predicted was over 10\%. This change was observed even when controlling for the day of the week and week of the year.

Results from the six countries where ChatGPT was unavailable also showed an increase in page views and visitor numbers in the period after release. Given this increase cannot be a result of ChatGPT, it may initially appear that the increases observed across the different languages was merely the result of changes in long-term engagement patterns across Wikipedia as a whole. However, what is notable is that the increases for languages from countries where ChatGPT was unavailable were consistently greater than for those languages from countries where it was. This may suggest that the release of the tool was associated with a lower increase in engagement within those particular Wikipedia language editions.

On the one hand, this would not be entirely surprising. The phenomenon of interdependence between Wikipedia and other Web services is well documented, particularly search engines \cite{mcmahon2017substantial,piccardi2021value,vincent2021deeper}. At this time, there is limited published research which demonstrates how and why users have been engaging with ChatGPT, but early indications would suggest users are turning to it in place of other information gathering tools such as search engines \cite{karunaratne2023new,taecharungroj2023can}. Indeed, question answering, search and recommendation are key functionalities of large language models identified in within the literature \cite{chang2023survey}. Unlike search engines, ChatGPT does not have a clear capacity to drive users to Wikipedia. It does not typically link to sources unless asked to \cite{wu2023towards} and even sometimes fabricates sources \cite{mcgowan2023chatgpt,zuccon2023chatgpt}. Even were it to link to sources, evidence from existing knowledge sharing systems suggests that any relationship is extremely one way with little -- if any -- traffic coming to Wikipedia from comments and answers that cite articles and pages \cite{vincent2018examining}. 

Conversely, it is essential to highlight that the languages we explored shared very different linguistic, geographic and cultural backgrounds. Smaller resourced Wikipedia language platforms may be more significantly impacted by the activities of bots \cite{chelsy2019detecting} and we were unable to select a range of community or resource sizes when selecting the six languages where ChatGPT was not available. By necessity, many of these communities are very small and although we tried to limit the impact of bots by requesting only contributions from users, there is no guarantee this would successfully filter out all bots. It is also possible that any observed effects are not platform-specific and may have been impacted by changes in the larger language communities. After all, prior research has demonstrated that more highly-resourced languages can be more influential within Wikipedia as changes to those languages may be more likely to propagate through to other language editions and are observed to propagate at higher speeds \cite{valentim2021tracking}. 

\subsection{Impact on Edits}

Any impact on editing behaviours was more limited. While we saw substantial changes in Arabic and Urdu -- and a weakly significant ~10\% fall in Swahili -- half of the languages we analysed did not record a statistically significant trend. We note that there are two factors that may diminish or even obfuscate any impact of ChatGPT and similar tools on edits. Firstly, prior work has outlined the importance of stigmergy where Wikipedia editors are encouraged to participate when becoming aware of -- or otherwise observing -- traces of other editors' activity \cite{zheng2023stigmergy}. In essence, edits can be self-multiplying as editors are drawn to build on, correct, replace or even remove other volunteers' contributions. It should also be emphasised that many Wikipedia edits are likely to be performed by bots across a whole host of roles although evidence suggests the number of bots has diminished significantly over time \cite{zheng2019roles}. In some languages such as Cebuano, the majority of pages are bot generated and user-generated content is lacking despite the language having a large number of articles that might warrant its classification as highly resourced \cite{anderson2017there}. Assuming bots are automated or semi-automated, we would not expect an immediate impact in edit numbers caused by those bots. 

Additionally, while a reduction in engagement has been seen in Stack Overflow and Stack Exchange \cite{del2023large,sanatizadeh2023information}, no such effect has been observed in other knowledge exchange communities such as Reddit \cite{burtch2023consequences}, something which may be linked to the social connections and relationships users form \cite{burtch2024generative}. While Wikipedia editors do form acquaintances through their discussion interactions \cite{jankowski2016verifying}, whether these are sufficient to influence the drive to edit and whether this drive would be sufficient to overcome any negative impact from ChatGPT is an area for future work. Nevertheless, we see no evidence of any change in edit and editor numbers that could be associated with the release and availability of ChatGPT.

\subsection{Limitations}

We recognise a number of limitations in our approach. Firstly, we have divided countries into groups based on the availability of ChatGPT, but we recognise there will be expatriate users, individuals using VPNs and potentially multilingual users who may have been able to circumvent these restrictions. Secondly, we have not accounted for the popularity of ChatGPT in a given country, in part due to the large number of countries involved in our analysis and the limited availability of data around the use of ChatGPT at the time this was written. We believe this is an important area for further work. We similarly were unable to account for the release of other LLMs and competitors to ChatGPT, although we note ChatGPT is likely to be the most popular. Thirdly and perhaps most significantly, although we accounted for seasonal effects in a given language, we were unable to account for cultural and national events that may have influenced the data. This was due largely to the significant number of countries where some languages were spoken (e.g., Arabic and English) and the large number of cultural or historic effects that may therefore arise. Our analysis may have been affected by one-off events, but also longer-term periodic events such as elections. Correcting for such events and exploring any effect they may have had is an important area for further study. 

\subsection{Future Work}

While our findings may not be conclusive, they nonetheless raise interesting questions about the impact LLMs and other AI tools may have on the wider Web infrastructure. Where evidence from Stack Overflow and Stack Exchange found a rapid and significant drop in engagement following the launch of ChatGPT, we find no such drop. Given the suggestion that highly social communities such as Reddit have been less impacted by LLMs, we question whether the communities within different language editions of Wikipedia may also be less impacted. In particular, we note interesting questions around whether articles with or fewer editors and articles with more or less discussion on their article talk pages demonstrate any difference in viewing and editing behaviours.

Beyond Wikipedia, however, our findings inevitably raise the question of how other large-scale knowledge sharing and collaborative systems may be impacted by the rise of large language models. We also note questions around how the domain of a service or article may influence whether it is impacted by large language models. For example, while the existing literature suggests that popular topics are language-independent, there nonetheless exist differences particularly surrounding cultural articles which will inevitably lead to variations between individual language editions of Wikipedia. Perhaps the largest opportunity for future work, however is the question of whether these shorter-term trends will hold in the longer term. 

\section{Conclusion}

In this paper, we analysed page visit, visitor, edit and editor numbers in twelve language editions of Wikipedia. We compared these metrics before and after the 30th of November 2022 when ChatGPT released and developed a panel regression model to better understand and quantify any differences. Our findings suggest an increase in page visits and visitor numbers that occurred across languages regardless of whether ChatGPT was available or not, although the observed increase was generally smaller in languages from countries where it was available. Conversely, we found little evidence of any impact for edits and editor numbers. We conclude any impact has been limited and while it may have led to lower growth in engagement within the territories where it is available, there has been no significant drop in usage or editing behaviours.

\section{Significance Statement}

Wikipedia is the world's largest encyclopedia, relying on collective intelligence from millions of users around the world to develop and maintain articles. In November 2022, the generative AI chat-bot ChatGPT released to the public and it has since grown rapidly to be one of the most popular services on the Web. While it is still relatively novel, early indications suggest that the release of ChatGPT has negatively impacted engagement in Web-based collective intelligence question-answering services such as StackOverflow. ChatGPT may fill some of the niches that Wikipedia currently occupies including fact finding and question-answering and some Wikipedia editors have raised the concern that the two platforms may end up in competition with one another.

To explore this, we analyse page-viewing and article editing activity within twelve language editions of Wikipedia. Six of these languages were predominantly spoken in countries where ChatGPT is available, while the other six were spoken in countries where it is unavailable or banned. Our analysis finds no evidence for an impact on edits or editor numbers, but we find some suggestions that page views and viewer numbers may have grown less in languages where ChatGPT was available than those where it was not. 

We posit that edits may have been unaffected as a more social, community-driven activity than anonymous page viewing. Our findings contribute to the emerging body of evidence of how ChatGPT is influencing collective intelligence tools although we caution that further research is needed to better understand any relationship between the two platforms.

\theendnotes


\end{document}